# Leveraging Novel Ensemble Learning Techniques and Landsat Multispectral Data for Estimating Olive Yields in Tunisia

Mohamed Kefi[1,a], Tien Dat Pham [2,+, a], Thin Nguyen[3], Mark G. Tjoelker [2], Viola Devasirvatham [2], and Kenichi Kashiwagi [4]

1. Water Research and Technologies Centre of Borj Cedria (CERTE), Soliman, Tunisia
2. Hawkesbury Institute for the Environment, Western Sydney University, Locked Bag 1797, Penrith, New South Wales 2751, Australia.
3. Applied Artificial Intelligence Institute (A2I2), Deakin University, Geelong Waurn Ponds Campus, 75 Pigdons Road, Waurn Ponds, Victoria 3216, Australia
4. Alliance for Research on the Mediterranean and North Africa (ARENA), The University of Tsukuba, Tsukuba, Ibaraki, Japan

[+]Correspondence: Dat.Pham@westernsydney.edu.au

[a] Authors contribute equally to this work.

## Abstract

Olive production is an important tree crop in Mediterranean climates. However, olive yield varies significantly due to climate change effects. Accurately estimating yield using remote sensing and machine learning remains a complex challenge. In this study, we developed a streamlined pipeline for olive yield estimation in the Kairouan and Sousse governorates of Tunisia. We extracted features from multispectral reflectance bands, vegetation indices derived from Landsat-8 OLI and Landsat-9 OLI-2 satellite imagery, along with digital elevation model (DEM) data. These spatial features were combined with ground-based field survey data to form a structured tabular dataset. We then developed an automated ensemble learning framework—implemented using AutoGluon—to train and evaluate multiple machine learning models, select optimal combinations through stacking, and generate robust yield predictions using five-fold cross-validation. The results demonstrate strong predictive performance from both sensors, with Landsat-8 OLI achieving $R^2$ = 0.8635 and RMSE = 1.17 tons ha$^{-1}$, and Landsat-9 OLI-2 achieving $R^2$ = 0.8378 and RMSE = 1.32 tons ha$^{-1}$. This study highlights a scalable, cost-effective, and accurate method for olive yield estimation, with potential applicability across diverse agricultural regions globally.

## 1. Introduction

In the Mediterranean region, the production of olives and olive oil constitutes an essential component of national economies. Olives (*Olea europaea* L.) are among the oldest cultivated crops and are primarily grown in Mediterranean countries (Carrion et al., 2010). The region's climate has significantly shaped its agricultural practices, which largely focus on vines, cereals, and olives (Ramos et al., 2025). However, climate change is expected to impact different regions and crops in varying ways. Arid and semi-arid areas, in particular, are projected to experience reduced precipitation and higher temperatures, leading to crop yield losses (FAO, 2018). Numerous factors influence the annual variation in olive oil production, including meteorological conditions such as temperature and precipitation fluctuations, economic dynamics, and agricultural practices like the adoption of irrigation (Fraga et al., 2021; Orlandi et al., 2020; Oteros et al., 2014; Kamiyama et al, 2021; Tous et al., 2010). Numerous crops in the Mediterranean areas are affected by climate change such as droughts and water scarcity (Fraga et al., 2022). Consequently, climate change is poised to have a significant impact on the economics of olive cultivation throughout the region (Ponti et al., 2014).

Tunisia, like many Mediterranean countries, is recognised for its production of olives and olive oil. In Tunisia, the olive sector has not only economic importance but also significant social and cultural value. Olive production fluctuates greatly from year to year due to the alternate bearing cycle of olive trees, characterized by low-yield "off" years followed by high-yield "on" years (Fichtner et al., 2018), as well as the influence of irregular precipitation and extreme weather conditions. Despite these challenges, Tunisia remains one of the world's leading producers and exporters of olive oil. Between the 2001/2002 and 2023/2024 seasons, Tunisia recorded an average annual olive oil production of 192,500 tons and average exports of approximately 152,000 tons (IOC, 2024). Olive oil exports play a significant role in the Tunisian economy and significantly contribute to the agricultural trade balance.

In this context, it is crucial for countries to monitor and evaluate olive and crop yields to develop and implement sustainable strategies. To maintain production levels and improve agricultural productivity, many countries support the agriculture sector by adopting advanced methods and technologies. Crop yield is influenced by multiple factors, including cultivation practices, climatic conditions, plant health, crop quality, and soil characteristics (Modi et al., 2022). Policymakers often aim to increase yields to keep food prices low and avoid the expansion of cropland into natural ecosystems (Lobell et al., 2009).

However, due to spatial variability in soil and climate conditions, accurately measuring and evaluating yield potential and yield gaps is challenging. A primary objective of yield prediction is to reduce uncertainty for farmers and provide knowledge to inform practices that support increased yield potential and avoid production shortages (Lobell et al., 2009). The accuracy of crop yield prediction depends on overcoming challenges related to data quality, feature selection, and environmental variability (Abu Jabed et al., 2024). In the past, agricultural forecasting depended on simple statistical models or expert judgment (Addanki et al., 2024). At present various innovative tools and technologies are available for production forecasting to narrow the yield gap and improve prediction accuracy.

Optical remote sensing data has proven its ability to model drought and identify olive-growing farms (Kefi et al., 2022; Kefi et al., 2016). Machine learning (ML) and deep learning (DL) when applied to remote sensing data enable solutions to complicated problems in various fields (Pham et al., 2023). ML models have been used to detect pest presence in olive and vineyard crops (Rodríguez-Díaz et al., 2024; Shawon et al., 2025). Addanki et al. (2024) applied ML and DL algorithms for crop prediction in India using the extreme gradient boosting (XGBoost) technique. Artificial Intelligence (AI) is proving invaluable in agriculture, offering the potential to predict key variables in near real-time to enable yield optimisation at minimal cost (Marwaha et al., 2023). Additionally, multispectral remote sensing has revolutionised the way agricultural data are collected and processed, playing a vital role in assessing olive cultivation (Anastasiou et al., 2023). To date, various efforts have been made to estimate olive yield using optical remote sensing data (Anastasiou et al., 2023; Marwaha et al., 2023; Rodríguez-Díaz et al., 2024; Addanki et al, 2024; Shawon et al., 2025). However, prediction performance has remained relatively low, with $R^2$ values less than 0.65. Thus, the main objective of this research is to improve the accuracy of yield prediction by developing novel ensemble learning techniques that integrate spectral reflectance and vegetation indices derived from multispectral remote sensors using Landsat imagery. This study also aims to compare the performance of the newly launched Landsat-9 OLI-2 sensor with that of Landsat-8 OLI for predicting olive yield using advanced ML technologies. Our research aims to contribute to the development of robust and reliable tools that can provide valuable insights for decision-makers and farmers.

## 2. Materials and Methods

### 2.1. Description of Study Areas

This study was carried out in olive-growing farms in the Tunisian governorates of Kairouan and Sousse. Kairouan is located in the central region of the country, while Sousse is a coastal area. In both prefectures, the olive-growing sector is well-developed and has a significant social and economic impact on local communities. The region is characterized by a semi-arid climate with moderate winters and highly variable precipitation. The annual average rainfall slightly exceeds 300 mm, and the average annual temperatures ranges between 16°C and 20°C. The highest temperatures are typically recorded in July and August. Like many other parts of Tunisia, Kairouan frequently experiences droughts (Kefi et al., 2022).

Tunisia is home to several olive cultivars, but the two main ones are Chetoui, which is primarily grown in the north, and Chemlali, which is widely cultivated in the central, coastal, and southern regions. Climate, cultivar type, and water availability are key factors influencing olive yield in Tunisia (Kefi et al., 2016). Based on production levels, the study selected Bouhajla, Nasrallah, Menzel Mhiri, and Chrarda in Kairouan, and Bouficha, Kalaa Kebira, and Msaken in Sousse. Figure 1 illustrates the locations of these selected study areas used for the analysis and monitoring of olive farm production.

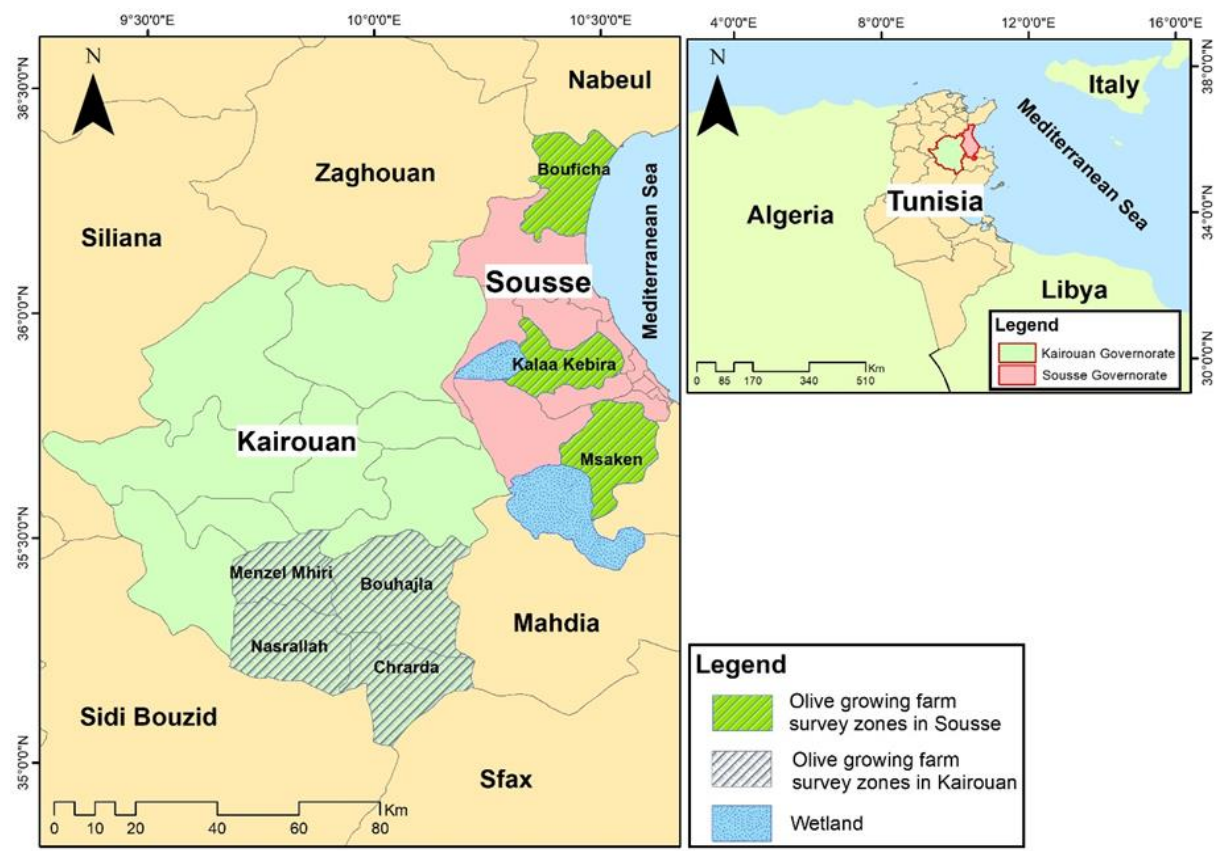

**Figure 1.** Location map of the study areas

## 2.2. Materials
### 2.2.1. Satellite Data Acquisition

Landsat surface reflectance (SR) data obtained through Earth Explorer (https://earthexplorer.usgs.gov/) was used to map vegetation dynamics in the study area (Table 1). We used Collection 2 products, which were atmospherically corrected SR data with a single-channel algorithm developed by the National Aeronautics and Space Administration (NASA) Jet Propulsion Laboratory (JPL). Landsat-8 OLI and Landsat-9 OLI-2 SR data used in the current study (Table 1) were acquired for the same period of the field survey. To minimize the effects of cloud, Landsat datasets with less than 5% cloud coverage, acquired in December 2022 and January 2023, coinciding with the dates of field surveys, were used for the analysis (Table 1). The Advanced Land Observing Satellite Digital Surface Model (DSM) at a horizontal resolution of 30 meters within the study area was used as an input feature for developing ML models https://www.eorc.jaxa.jp/ALOS/en/dataset/aw3d30/index.htm.

**Table 1.** Landsat data acquisition in the study area

| Sensor | Spatial resolution (m) | Image_ID | Cloud Cover (%) | Band used |
|---|---|---|---|---|
| Landsat-8 OLI | 30 | LC08_L2SP_191035_20221227_20230104 | 2.93 | Coastal, Blue, Green, Red, Near-Infrared (NIR), Mid-Infrared (MIR), Shortwave infrared 1 (SWIR 1), Shortwave infrared 2 (SWIR 2) |
| | | LC08_L2SP_191036_20221227_20230104 | 0.23 | |
| Landsat-9 OLI-2 | 30 | LC09_L2SP_191035_20230104_20230314 | 3.43 | |
| | | LC09_L2SP_191036_20230104_20230314 | 0.11 | |

### 2.2.2. Field Survey Data

The survey was conducted at the level of olive-growing farms through four cities in Kairouan and three cities in Sousse. We conducted surveys during December 2022 and January 2023 with local stakeholders from each study area. A face-to-face approach was used to collect data on olive production. Interviews were conducted with local olive producers, who were randomly selected based on specific criteria, including farm size and irrigated or rainfed farms.

A specific questionnaire was used to gather relevant information about olive production in the field. During the interview, we informed the farmers that the data collected would be used solely for research purpose and their personal information would not be disclosed. During the survey, ground truth points (GTPs) were recorded using GPS (Global Positioning System) to accurately locate the visited olive farms. The questionnaire covered several aspects such as farm characteristics, olive production by tree and by hectare, agricultural practices, and production costs. A total of 192 olive farms were sampled for the analysis. The survey locations for each study area applied for this research are presented in Table 2.

**Table 2.** The locations of olive-growing farms

| Governorate | City | Farms |
|---|---|---|
| Kairouan | Bouhajla | 32 |
|  | Chrarda | 48 |
|  | Menzel Mhiri | 38 |
|  | Nasrallah | 5 |
| **Total** |  | **123** |
| Sousse | Bouficha | 30 |
|  | Kalaa Kebira | 36 |
|  | Msaken | 3 |
| **Total** |  | **69** |

### 2.3. Methods

We propose a novel multi-layer stacked ensemble regression model for estimating olive yield using Landsat-8 OLI and Landsat-9 OLI-2. Our model development consists of the following steps: (1) pre-processing satellite optical data (7 spectral bands), computing vegetation indices, and incorporating a digital elevation model (DEM); (2) integrating the field survey data and extracting features into a tabular format; (3) testing the model using AutoGluon with eight ensemble learning algorithms; (4) evaluating and comparing the performance of the Landsat sensors using 5-fold cross-validation. Figure 2 illustrates the framework for estimating olive yield from Landsat imagery in this study. The flowchart illustrates an end-to-end pipeline for olive yield estimation. The raw Landsat-8 OLI and Landsat-9 OLI-2 data are pre-processed to extract multispectral features and vegetation indices, which are then combined with field survey data to create a tabular dataset. This dataset serves as input to a unified machine learning framework that performs model training, automated ensemble selection, and cross-validation to generate accurate yield predictions.

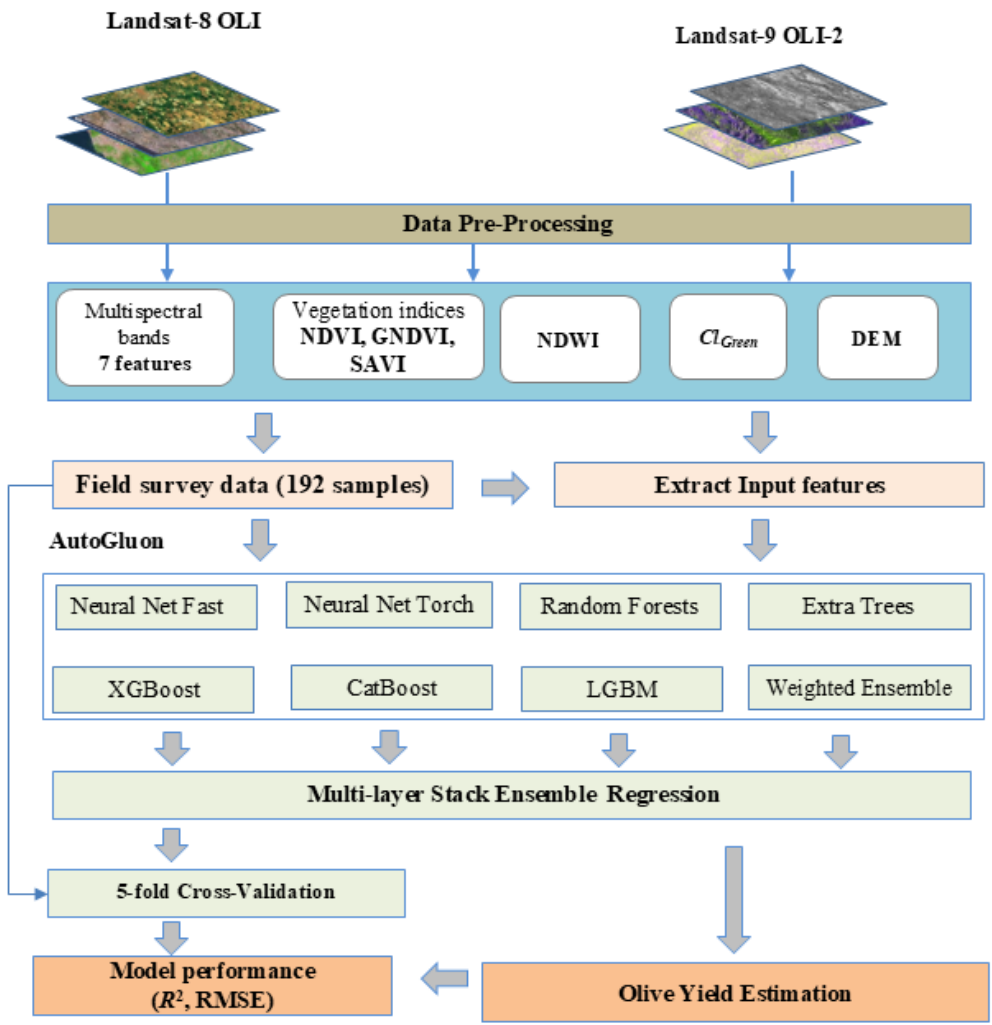

**Figure 2.** Flowchart for developing novel ensemble learning model using Landsat data

*2.3.1. Satellite image processing*

Before performing image analysis, preprocessing was applied to Landsat-8 OLI and Landsat-9 OLI-2 data using ArcGIS Pro V3.2 tools. To accurately identify landscape features, the first step involved creating a single raster from multiple spectral bands. Seven multispectral bands (B1, B2, B3, B4, B5, B6, and B7) from Landsat imagery with a spatial resolution of 30 m, were used to generate band composites. To cover the entire study area, we created a mosaic of two

Landsat scenes using the Mosaic Dataset tool in ArcGIS Pro V3.2. Histogram matching was applied for relative radiometric normalization to ensure consistent radiometric properties, such as brightness, contrast, and reflectance values (Nyamtseren et al., 2025). The dataset was then clipped to define the research study area. All raster data used in the analysis were georeferenced to the Universal Transverse Mercator (UTM) coordinate system, WGS84 datum, zone 32N.

To monitor olive production and identify land-cover features, various vegetation and water indices were computed from Landsat-8 OLI and Landsat-9 OLI-2 imagery. These included the Normalized Difference Vegetation Index (NDVI), Green Normalized Difference Vegetation Index (GNDVI), Soil-Adjusted Vegetation Index (SAVI), Chlorophyll Index Green ($CI_{green}$), and the Normalized Difference Water Index (NDWI), which are highly used in agriculture (Stateras and Kalivas, 2020; Moussaid et al., 2022; Marques et al., 2024).

The Normalized Difference Vegetation Index (NDVI) (Rouse et al., 1974) (Eq. 1 ) was used to identify olive-growing farms within the study area. The Green Normalized Difference Vegetation (GNDVI) (Gitelson et al.,1996) (Eq. 2) derived from the NIR and Green bands, was also computed as it indicates photosynthetic activity and is more sensitive to chlorophyll concentration. To minimize the influence of soil brightness on vegetation indices, the Soil-Adjusted Vegetation Index (SAVI) (Eq. 3) was applied (Huete, 1988).

The Normalized Difference Water Index (NDWI) (Gao, 1996) (Eq. 4) was also computed in the analysis as it is highly correlated with canopy water content. Finally, $Cl_{green}$ index (Yin et al., 2016) (Eq. 5) derived from NIR and the Green surface reflectance band, was computed to further support olive crop identification.

The equations of spectral indices are listed below:

$$\text{NDVI} = \frac{\rho_{nir} - \rho_{red}}{\rho_{nir} + \rho_{red}} \qquad (1)$$

$$\text{GNDVI} = \frac{\rho_{nir} - \rho_{green}}{\rho_{nir} + \rho_{green}} \qquad (2)$$

$$\text{SAVI} = (1 + L) \times \frac{\rho_{nir} - \rho_{red}}{\rho_{nir} + \rho_{red} + L} \qquad (3)$$

$L = 0.5$ in most conditions

$$\text{NDWI} = \frac{\rho_{swir1} - \rho_{nir}}{\rho_{swir1} + \rho_{nir}} \qquad (4)$$

$$Cl_{green} = \frac{\rho_{nir}}{\rho_{green}} - 1 \qquad (5)$$

where $\rho_{green}$, $\rho_{red}$, and $\rho_{nir}$ are the surface reflectance at the green wavelength (band 3, 560 nm), red wavelength (band 4, 655 nm), and near-infrared wavelength (band 5, 865 nm) bands, respectively.

All 13 features were computed for both Landsat-8 OLI and Landsat-9 OLI-2 datasets and transformed into Tabular format (.csv) beforehand for the regression task.

*2.3.2. Development of Advanced Machine Learning Techniques*

In this study, we developed a novel machine learning model by incorporating seven multispectral bands and four vegetation indices including NDVI, GNDVI, SAVI, and *Cl*green along with NDWI, and a DEM into the AutoGluon framework to predict olive yield. AutoGluon, initially introduced by Amazon Co. Ltd., provides a platform for the automated design and implementation of ensemble learning models across various applications (Erickson et al., 2020). Currently, AutoGluon is among the most widely used ensemble learning architectures, employing various decision tree-based ensemble methods. It effectively supports both classification and regression tasks within supervised learning domains. AutoGluon also automates hyperparameter tuning and model architecture search, enabling the automatic design of optimal ensemble models (Qi et al., 2021).

In our work, we employed the AutoGluon framework to import the tabular .csv files. Its tabular prediction module includes eight customized models, such as two neural network models—Neural Net Fast and Neural Net Torch—Random Forest, Extra Trees, three boosting algorithms (CatBoost, Extreme Gradient Boosting (XGBoost), Light Gradient Boosting Machine (LGBM), and a weighted ensemble model (Erickson et al., 2020).

Model development and evaluation were conducted using Python 3.12 in a Jupiter Notebook environment. The machine learning models were implemented using libraries from Scikit-learn [39], and AutoGluon, available at https://auto.gluon.ai/

*2.3.3. Accuracy Assessment*

A total of 192 samples were used to evaluate the model performance using the standard metrics in the Scikit-learn library (Pedregosa et al., 2011). We tested the ML models with all 13 features (predictor variables) using a 5-fold cross-validation (CV) with automated hyperparameter tuning. We compared the highest predictive model based on the lowest root-mean-square error (RMSE, Eq. 6) and the highest coefficient of determination ($R^2$, Eq. 7).

$$\text{RMSE} = \sqrt{\sum_{1}^{n} \frac{(ye_i - ym_i)^2}{n}}, \qquad (6)$$

$$R^2 = \frac{\sum_{i=1}^{n}(ye_i - \overline{ye})(ym_i - \overline{ym})}{\sqrt{\sum_{i=1}^{n}(ye_i - \overline{ye})^2 (ym_i - \overline{ym})^2}}, \qquad (7)$$

Where $y_{ei}$ is the estimated olive yield value from the ML model, is the actual olive yield value obtained from the field survey, $n$ is the total number of sampling plots, $y_e$ and $y_m$ are the mean values of the estimated olive yield and the actual yield, respectively.

## 3. Results
### 3.1. Field survey results

During the survey, we selected 123 olive-growing farms in Kairouan and 69 in Sousse prefectures for analysis. The survey was conducted in December 2022 and January 2023, focusing on data related to the 2022/2023 production season (Table 3).

The average farm area observed in Kairouan is approximately 5.8 hectares, while in Sousse it is about 3.5 hectares. The survey revealed that the average yield in Kairouan is four times higher than in Sousse (Table 3).

**Table 3.** Overview of Survey data collected

| Prefecture | Sample size ($n$) | Average farm area (ha) | Number of irrigated olive farms | Average age of trees | Olive Tree density (trees ha$^{-1}$) | Average yield (Tons ha$^{-1}$) | Average Elevation (m) |
|---|---|---|---|---|---|---|---|
| Kairouan | 123 | 5.8 | 120 | 53 | 44 | 4.6 | 46.1 |
| Sousse | 69 | 3.5 | 48 | 84 | 61 | 1.1 | 134.7 |

In our sample, a vast majority—87% of olive farms—are smaller than 10 hectares. As shown in Table 4, 66% of farms are smaller than 5 hectares, 21% range between 5 and 10 hectares, and only 13% exceed 10 hectares. Additionally, orchard density and tree age may influence olive production. The average tree age is 53 years in Kairouan and 84 years in Sousse, reflecting the presence of older orchards in both regions. Tree density is 44 trees per hectare in Kairouan and 61 trees per hectare in Sousse. Based on GPS field data collected, the average elevation in Kairouan is approximately 135 m above sea level, while in Sousse, it is around 46 m.

**Table 4.** Characteristics of Olive growing farms in the study sites

| Prefecture | Number of olive-growing farms | | | |
|---|---|---|---|---|
| | < 5 ha | 5 ha $\geq$ area < 10 ha | $\geq$10ha | Total |
| Kairouan | 71 | 31 | 21 | 123 |
| Sousse | 56 | 9 | 4 | 69 |
| Total | 127 | 40 | 25 | 192 |

*3.2. Machine learning modelling, performance, and comparison*

As shown in Table 5, Landsat-8 OLI yielded slightly higher performance using 5-fold CV, with a mean $R^2$ of 0.8635 and a mean RMSE of 1.167 tons ha$^{-1}$, while Landsat-9 OLI-2 produced satisfactory results, with a mean $R^2$ of 0.8378 and a mean RMSE of 1.322 tons ha$^{-1}$. The highest $R^2$ values were archived at 0.9201 for Landsat-8 and 0.8778 for Landsat-9, while the lowest $R^2$ values were 0.7989 and 0.7544 for Lansat-8 and Landsat-9, respectively (Table 5).

**Table 5.** Model performance and comparison using the advanced AutoGluon framework

| | Landsat-8 OLI | | Landsat-9 OLI-2 | |
|---|---|---|---|---|
| Fold | $R^2$ | RMSE (tons ha $^{-1}$) | $R^2$ | RMSE (tons ha $^{-1}$) |
| 1 | 0.8035 | 1.251 | 0.7544 | 1.793 |
| 2 | 0.8985 | 1.227 | 0.8661 | 1.358 |
| 3 | 0.9201 | 0.938 | 0.8208 | 1.464 |
| 4 | 0.8967 | 0.735 | 0.8778 | 0.853 |
| 5 | 0.7989 | 1.683 | 0.8698 | 1.139 |
| **Average** | **0.8635** | **1.167** | **0.8378** | **1.322** |

Figure 3 shows scatterplots of predicted versus actual olive yields across 5-fold CV. The Landsat-8 sensor yielded higher predictive capability and better goodness-of-fit. Both two sensors were able to predict olive yield with precise accuracy, up to 10 tons ha $^{-1}$. However, the scatter plots indicate data saturation in the Landsat imagery at yield values exceeding 15 tons ha $^{-1}$.

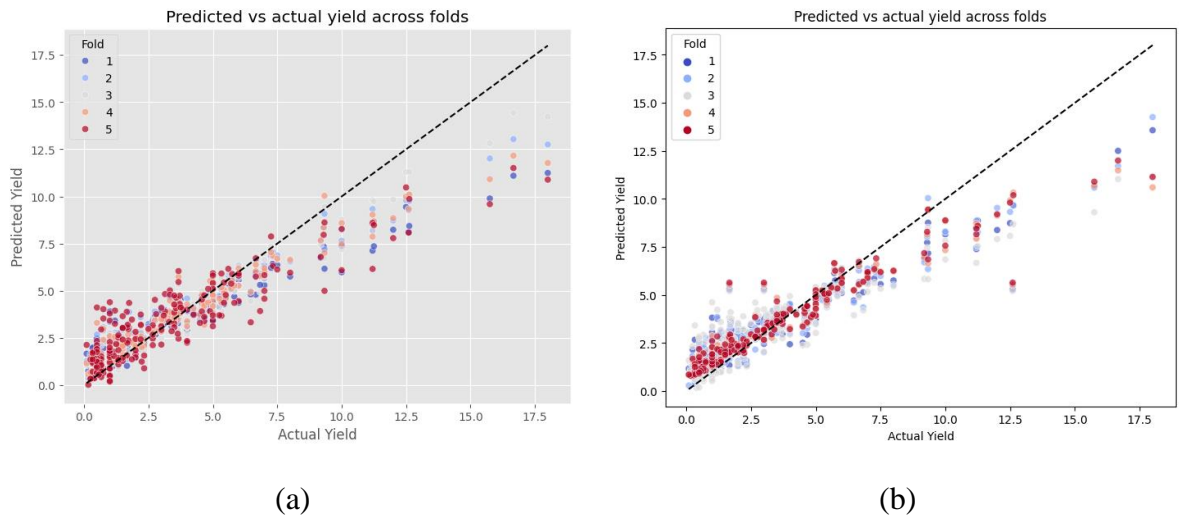

(a)          (b)

**Figure 3.** Scatter plots of estimated versus actual olive yields across a 5-fold CV procedure.
(a) Landsat-8 OLI, (b) Landsat-9 OLI-2. The 1:1 lines are shown.

*3.3. Variable importance*

Figure 4 presents the variable importance for predicting olive yield in the study areas using the Landsat-8 OLI and Landsat-9 OLI-2 datasets. A similar trend is observed across both datasets. The most important feature across the 5-fold CV is Band 13, representing the DEM, with an average importance of more than 50%. Band 5 (NIR), band 7 (SWIR), NDVI (Band 8), and NDWI (Band 11) are also significant features. In contrast, Band 12 ($Cl_{Green}$) is among the least important, followed by the SAVI (Band 10).

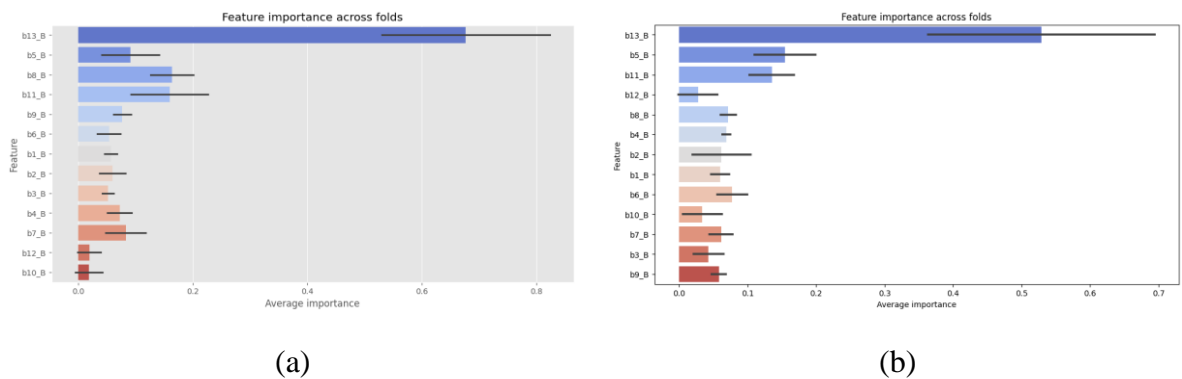

(a)          (b)

**Figure 4.** Variable importance of band features for predicting olive yield
(a) Landsat-8 OLI, (b) Landsat-9 OLI-2

## 4. Discussion
*4.1. Olive production and drivers*

In both study areas, Kairouan and Sousse, olive yields are relatively low, which can be attributed to several factors (Table 3). The predominant cultivar grown is *Chemlali*, which is well adapted to drought conditions and climate variability. However, it is also characterised by an alternating bearing cycle, which impacts yield consistency among farms. Farm size is another significant factor due to the subdivision of agricultural land in Tunisia, approximately 72% of olive farms are smaller than 10 hectares (FAO, 2015).

Irrigation also plays a crucial role in enhancing productivity. As shown in Table 4, 98% of surveyed farms in Kairouan are irrigated, compared to only about 70% in Sousse. Expanding irrigation to currently non-irrigated farms could significantly improve olive production (Kashiwagi et al., 2013; Kamiyama et al., 2016). The survey revealed that yields are higher in Kairouan than in the Sousse prefecture, likely due to differences in tree age, elevation, and irrigation practices.

*4.2. Advances in ensemble-learning decision trees in estimating olive yield*

Accurate yield estimation is critical for improving crop management and reducing costs. Various attempts have been made using different optical sensors and UAVs-based images for yield estimation, as recently reported by Marques et al. (2024). Over the past 16 years, a wide array of remote sensing techniques and platforms has been applied to olive and other orchard crops (Anastasiou et al., 2023).

Stateras and Kalivas (2020) employed high spatial resolution UAV data to predict olive yield in Greece, achieving an $R^2$ of 0.6. However, single machine learning algorithms such as Support Vector Machine (SVM), Gaussian Process (GP), artificial neural networks (ANN) as well ensemble models like Random Forest (RF), CatBoost, XGBoost have shown limited predictive performance, often with $R^2$ values below 0.65 for olive and other orchard crops such as citrus (Stateras and Kalivas, 2020; Cubillas et al., 2022; Moussaid et al., 2022; Ramos et al., 2025). In contrast, our proposed model performed well and outperformed these previous approaches by leveraging freely available Landsat data, achieving $R^2$ values exceeding 0.83 (Table 5).

Variable importance analysis in Fig. 4 identified elevation as the most influential predictor, with higher elevations associated with lower yields. This finding is strongly supported by Table 3 and is consistent with results reported by Stateras and Kalivas (2020). The likely reason is climatic influence — higher altitudes tend to be more exposed to wind, have reduced water availability, and are at greater risk of frost, which can damage flowers and reduce yield (Leauthaud et al., 2022). Vegetation indices such as NDVI and NDWI were also key features for predicting olive yield in Mediterranean regions and strongly align with recent studies (Moussaid et al., 2022; Ramos et al., 2025). Amongst the spectral reflectance bands, the near-infrared (NIR) wavelength is the most important, followed by the shortwave infrared (SWIR). The results are consistent with previous studies for crop yield monitoring using optical sensors (Fieuzal et al., 2017; Moussaid et al., 2022).

*4.2. Limitations and sensor saturation*

Data saturation is a common issue when using optical remote sensing sensors for crop yield prediction. Optical sensors tend to overpredict yield at lower values and underpredict it at very high values. For instance, Landsat sensors often become saturated at yields exceeding 15 tons per hectare (Fig. 3), while Sentinel-2 provides a wider range of wavelengths and more spectral bands, thus supporting higher saturation thresholds (Fieuzal et al., 2017; Marques et al., 2024).

To overcome the data saturation limitations of individual sensors, fusing multisensor data—such as combining optical imagery with synthetic aperture radar (SAR) or LiDAR—offers a promising alternative for improving crop yield estimates. Fieuzal et al. (2017) successfully combined optical and radar satellite data to accurately estimate early-stage corn yields. Future research should focus on integrating multisource and multisensor data to enhance the accuracy of crop yield estimations.

With the rapid advancement of computer vision and artificial intelligence, the development of new algorithms capable of handling small datasets—such as tabular foundation models (Hollmann et al., 2025), offers promising prospects for accurate crop yield prediction in the near future.

## 5. Concluding remarks

The development of novel machine learning models for olive yield prediction holds significant potential to support farmers in optimising cultivation practices. In this study, we developed a new ML model that integrates spectral indices, band reflectance, and a DEM within the AutoGluon framework using eight ensemble learning algorithms. Our results demonstrated precise yield estimates using Landsat imagery. Landsat-8 OLI achieved slightly higher performance, with an $R^2$ of 0.8635 while Landsat-9 OLI-2 yielded an $R^2$ of 0.8378. We also found that elevation was the most important factor for estimating olive yield, with higher elevations associated with lower yields. NIR and SWIR wavelengths were identified as the most important spectral features. NDVI and NDWI were significant vegetation indices for olive yield estimation. Future research on olive yield prediction should explore the use of other multispectral sensors that capture longer wavelengths and offer additional spectral bands, particularly in Mediterranean climate zones and regions with diverse geographic conditions. This would enhance the generalisability and scalability of the proposed method.

**Author Contributions:** Conceptualization, T.D.P. and M.K.; methodology, T.D.P.; T.N. and M.K.; software, T.D.P. and M.K.; validation, T.D.P; T.N. and M.K.; formal analysis, T.D.P.; M.K., T.N.; investigation, M.K.; resources, T.D.P.; M.K., T.N.; data curation, T.D.P.; M.K.; writing—original draft preparation, T.D.P., M.G.T. and M.K.; writing—review and editing, M.K., T.D.P., T.N., V.D., M.G.T. and K.K.; visualization, T.D.P., M.G.T., V.D., K.K and M.K.; supervision, M.G.T. and K.K.; project administration, M.K. and K.K; funding acquisition, K.K. All authors have read and agreed to the published version of the manuscript.

**Funding:** Data collection and survey were supported by the Japan Science for Promotion of Sciences (JSPS) Grant-in-Aid for Scientific Research, No. 20H03082 entitled: Development of olive industrial clusters towards transformation to decolonized structure by establishment of innovative value chain and were partially supported by JSPS Grant-in-Aid for Scientific Research, 24H00531 entitled: Empirical investigation of nudge for creation of innovative olive production cluster.

**Data Availability Statement:** Data are available upon request.

**Acknowledgments:** The authors would like to thank local authorities and local farmers for their generous assistance during the fieldwork and technical data process.

**Conflicts of Interest:** The authors declare no conflicts of interest.